\documentclass{PoS}

\usepackage[dvipdfm,dvips]{} 

\makeatletter
\def\simleq{\mathrel{\mathpalette\gl@align<}}
\def\simgeq{\mathrel{\mathpalette\gl@align>}}
\def\gl@align#1#2{\lower.6ex\vbox{\baselineskip\z@skip\lineskip\z@
     \ialign{$\m@th#1\hfill##\hfil$\crcr#2\crcr\sim\crcr}}}
\makeatother

\newcommand{\Luscher}{L\"uscher}

\title{%
First results of baryon interactions from lattice QCD with physical masses (1)
-- General overview and two-nucleon forces --
}

\ShortTitle{%
Baryon forces from lattice QCD with physical masses (1)
-- Overview and nuclear forces --
}

\author{
\speaker{Takumi~Doi},$^a$ 
Sinya~Aoki,$^{abc}$
Shinya~Gongyo,$^{ab}$
Tetsuo~Hatsuda,$^{ad}$
Yoichi~Ikeda,$^a$
Takashi~Inoue,$^{ae}$
Takumi~Iritani,$^{af}$
Noriyoshi~Ishii,$^{ag}$
Takaya~Miyamoto,$^{ab}$
Keiko~Murano,$^{ag}$
Hidekatsu~Nemura,$^{ac}$
and
Kenji~Sasaki$^{ac}$ \\
\llap{$^a$} Theoretical Research Division, Nishina Center, RIKEN, Wako 351-0198, Japan\\
\llap{$^b$} Yukawa Institute for Theoretical Physics, Kyoto University, Kyoto 606-8502, Japan\\
\llap{$^c$} Center for Computational Sciences, University of Tsukuba, Ibaraki 305-8571, Japan\\
\llap{$^d$} Kavli IPMU (WPI), The University of Tokyo, Chiba 277-8583, Japan\\
\llap{$^e$} Nihon University, College of Bioresource Sciences, Kanagawa 252-0880, Japan\\
\llap{$^f$} Department of Physics and Astronomy, Stony Brook University, Stony Brook, New York 11794-3800, USA\\
\llap{$^g$} Research Center for Nuclear Physics (RCNP), Osaka University, Osaka 567-0047, Japan\\
E-mail: \email{doi@ribf.riken.jp}}

\abstract{
We present the lattice QCD studies 
for baryon-baryon interactions 
for the first time with (almost) physical quark masses.
$N_f = 2+1$ gauge configurations are generated with
the Iwasaki gauge action and 
nonperturbatively ${\cal O}(a)$-improved Wilson quark action with stout smearing
on the lattice of $(96 a)^4 \simeq (8.2 {\rm fm})^4$ with $a \simeq 0.085$ fm,
where 
$m_\pi \simeq 146$ MeV and $m_K \simeq 525$ MeV.
Baryon forces are calculated from Nambu-Bethe-Salpeter (NBS) correlation functions
using the time-dependent HAL QCD method.
In this report, we first give the general overview 
of the theoretical frameworks
essential to the physical point calculation of baryon forces. 
We then present the numerical results for the two-nucleon 
central and tensor forces in $^3S_1$-$^3D_1$ coupled channel
and
the central force in $^1S_0$ channel.
In particular, a clear signal is obtained for the tensor force.
}

\FullConference{The 33rd International Symposium on Lattice Field Theory\\
		14 -18 July 2015\\
		Kobe International Conference Center, Kobe, Japan*}

\begin{document}

\vspace*{-17mm}
\section{Introduction}
\vspace*{-4mm}
\label{sec:intro}

Interactions among baryons (nuclear forces and hyperon forces) are
one of the most fundamental constituents in nuclear physics.
For nuclear forces, 
thanks to the extensive studies 
since the celebrated work by Yukawa 80 years ago~\cite{Yukawa:1935xg},
precision two-nucleon forces have been established.
They have been, however, obtained phenomenologically 
from experimental scattering phase shifts
and the relation to the underlying theory, Quantum Chromodynamics (QCD), is still hidden in a veil.
For hyperon forces,
even phenomenological information 
suffers from large uncertainties,
since scattering experiments with hyperon(s) are very difficult.
In the meantime,
the comprehensive determination of baryon forces
based on 
QCD
has become 
an urgent issue in not only nuclear physics but also astrophysics,
in the context of, e.g., the equation of state (EoS) of high dense matter.
%
%
Therefore, it is most desirable to 
make a systematic determination of baryon forces 
by the first-principles calculations of QCD, such as lattice QCD simulations.

In lattice QCD, baryon interactions have been traditionally studied
using the \Luscher's finite volume method~\cite{Luscher:1990ux}. 
(See, e.g., Refs.~\cite{Yamazaki:2011nd, Beane:2011iw, Berkowitz:2015eaa} for recent works.)
Recently, 
a novel theoretical framework, HAL QCD method, is proposed where
the interaction kernels (so-called ``potentials'') 
are determined from Nambu-Bethe-Salpeter (NBS) wave functions on a lattice~\cite{Ishii:2006ec}.
In particular, with the extension to the ``time-dependent'' HAL QCD method,
energy-independent (non-local) 
potentials can be extracted
without relying on the ground state saturation~\cite{HALQCD:2012aa}.
The method has been successfully applied to 
nuclear forces, hyperon forces including coupled channel systems,
three-nucleon forces and so on~\cite{Aoki:2012tk, HAL:recent}.
Obtained interactions are also used to predict the properties of 
medium-heavy nuclei, EoS of nuclear matter and mass-radius relation of neutron stars~\cite{Inoue:2013nfe}.

Albeit the significant progress in recent years,
existing lattice calculations for baryon interactions suffer from various systematic uncertainties,
where the most critical one is associated with unphysically heavy quark masses.
Under these circumstances, 
we have launched a new project 
under the HPCI (High Performance Computing Infrastructure) 
SPIRE (Strategic Program for Innovative REsearch) Field 5,
which aims at the lattice calculations of nuclear and hyperon potentials
with physically light quark masses on a large lattice volume,
exploiting the capability of the latest supercomputers such as Japanese flagship K computer.
In this paper, we present the latest status report for the 
baryon force calculations in this (on-going) project.
We first give general overview of the theoretical framework.
We then present numerical results for 
two-nucleon (2N) central and tensor forces 
in parity-even channel, namely, $^3S_1$-$^3D_1$ coupled channel and $^1S_0$ channel.
The results for hyperon forces obtained in the same lattice setup are presented
in Refs.~\cite{Ishii:lat2015}.
Detailed account on the generation of gauge configurations used in this study
is given in Ref.~\cite{Ukita:lat2015}.

\vspace*{-4mm}
\section{Formalism}
\vspace*{-3mm}
\label{sec:formalism}

We explain the HAL QCD method for the 2N system as an illustration.
We consider the (equal-time) NBS wave function in the center-of-mass frame,
%
$
\phi_W^{2N}(\vec{r}) \equiv 
1/Z_N \cdot
\langle 0 | N(\vec{r},0) N(\vec{0},0) | 2N, W \rangle_{\rm in} ,
$
%
where 
$N$ is the nucleon operator
with its wave-function renormalization factor $\sqrt{Z_N}$ 
and
$|2N, W \rangle_{\rm in}$ denotes the asymptotic in-state of the 2N system 
at the total energy of $W = 2\sqrt{k^2+m_N^2}$
with 
the relative momentum $k \equiv |\vec{k}|$,
and we consider the elastic region,  $W < W_{\rm th} = 2m_N + m_\pi$.
The NBS wave function may be extracted from the four-point correlation function
$
G^{2N} (\vec{r},t)
\equiv
\frac{1}{L^3}
\sum_{\vec{R}}
\langle 0 |
          (N(\vec{R}+\vec{r}) N (\vec{R}))(t)\
\overline{(N N)}(t=0)
| 0 \rangle .
$ 
%
%
%
The most important property of the NBS wave function is that
it has a desirable asymptotic behavior
at $r \equiv |\vec{r}| \rightarrow \infty$,
%
$
\phi_W^{2N} (\vec{r}) \propto 
\sin(kr-l\pi/2 + \delta_W^l) / (kr),
$
%
where 
$\delta_W^l$ is the scattering phase shift
with the orbital angular momentum $l$~\cite{Luscher:1990ux, Ishii:2006ec, Lin:2001ek}.
Exploiting this feature,
we define the (non-local) 2N potential, $U^{2N}(\vec{r},\vec{r}')$,
through the Schr\"odinger equation,
\begin{eqnarray}
%
(E_W^{2N} - H_0) \phi_W^{2N}(\vec{r})
= 
\int d\vec{r}' U^{2N}(\vec{r},\vec{r}') \phi_W^{2N}(\vec{r}')
%
\label{eq:Sch_2N:tindep}
\end{eqnarray}
where 
$H_0 = -\nabla^2/(2\mu)$ and
$E_W^{2N} = k^2/(2\mu)$ with the reduced mass $\mu = m_N/2$.
It is evident that
$U^{2N}(\vec{r},\vec{r}')$
is faithful to the phase shift by construction.
In addition,
it has been proven that 
one can construct $U^{2N}(\vec{r},\vec{r}')$
in an energy-independent way~\cite{Ishii:2006ec,Aoki:2012tk}.
Therefore, once $U^{2N}(\vec{r},\vec{r}')$ are obtained on a (finite volume) lattice,
we can determine the phase shifts at arbitrary energies below the inelastic threshold
by solving the Schr\"odinger equation with $U^{2N}(\vec{r},\vec{r}')$ in the infinite volume.

In the practical lattice calculations,
several additional developments are necessary,
in particular when calculations with physically light quark masses 
and larger volumes are of interest.
In the following, we present three crucial developments:
(1) extension to the ``time-dependent'' HAL QCD method
(2) extension to coupled channel formalism above inelastic threshold
and
(3) development in computational scheme called the unified contraction algorithm.

\subsection{Time-dependent HAL QCD method}
\label{subsec:t-dep-HAL}

In the original HAL QCD method described above,
it is necessary to isolate each energy eigenstate on a lattice
(most typical example is the so-called ground state saturation).
Unfortunately, 
this
is very difficult 
due to the existence of nearby elastic scattering states.
For instance, on a lattice with the spacial extent of $L = 8.2$ fm
and with physical quark masses
(which is the lattice setup in this study),
the first excited state of elastic 2N scattering state has
$\sim (2\pi/L)^2 / m_N \sim 25$ MeV excitation energy.
Correspondingly, ground state saturation 
requires 
$t \gg (25 {\rm MeV})^{-1} \sim {\cal O}(10) {\rm fm}$,
where $S/N$ 
may be
suppressed by a factor of 
$\exp[ -2(m_N - 3/2 m_\pi) t] \sim 10^{-32}$.

The time-dependent HAL QCD method~\cite{HALQCD:2012aa} is a suitable framework to avoid this problem.
By noting that 
$U^{2N}(\vec{r},\vec{r}')$ is energy-independent below $W_{\rm th}$,
one can show that 
the following ``time-dependent'' Schr\"odinger equation holds
even without the ground state saturation,
\begin{eqnarray}
\left( 
- \frac{\partial}{\partial t} 
+ \frac{1}{4m_N} \frac{\partial^2}{\partial t^2} 
- H_0
\right)
R^{2N}(\vec{r},t) 
=
\int d\vec{r}' U^{2N}(\vec{r},\vec{r}') R^{2N}(\vec{r}',t) ,
\label{eq:Sch_2N:tdep}
\end{eqnarray}
where
$R^{2N}(\vec{r},t)$ is the NBS correlation function given by 
$R^{2N}(\vec{r},t) \equiv G^{2N} (\vec{r},t) / \{G^N(t)\}^2$
with
$G^{N} (t)$ being
the single nucleon correlator.
It is still necessary to achieve ``elastic state saturation'' 
(suppression of contaminations from inelastic states),
but it can be fulfilled by much easier condition,
$t \gg (W_{\rm th} - W)^{-1} \sim m_\pi^{-1} \sim {\cal O}(1) {\rm fm}$.
This is in contrast to the \Luscher's method,
which inevitably relies on the ground state saturation.
(For detailed comparison between the \Luscher's method and the time-dependent HAL QCD method, 
see Refs.~\cite{Kurth:2013tua, Iritani:2015dhu}.)
%

\subsection{HAL QCD method for coupled channel systems}
\label{subsec:CC}

If we consider the hyperon forces, 
it is necessary to extend 
our framework so that it can be applied even above the inelastic threshold.
For instance, in the calculation of $\Lambda \Lambda$ interactions,
the inelastic threshold (the $N\Xi$ channel) is open only $\sim 30$ MeV above the
$\Lambda \Lambda$ threshold.
In addition, these coupled channel effects themselves are important subjects 
in hypernuclear physics.

The HAL QCD method can be extended to coupled channel systems
thanks to the energy-independence of (coupled channel) potentials~\cite{Aoki:2012tk}.
Considering the $\Lambda \Lambda$--$N \Xi$ coupled channel system as an example,
coupled channel potentials $U^c_{\ \ c'}(\vec{r},\vec{r}')$ can be extracted from the following equations 
in the original (time-independent) HAL QCD method,
\begin{eqnarray}
  \label{EQ:CoupledSE}
 \left( E_W^c  - H_0^{c} \right)
  \phi^{c}_{W}(\vec{r}) 
=
\sum_{c' = a,b}
\int d\vec{r}' U^c_{\ \ c'}(\vec{r},\vec{r}') \phi^{c'}_W(\vec{r}',t) 
\quad (c = a, b) ,
\end{eqnarray}
where $a, b$ denotes $\Lambda\Lambda$ and $N\Xi$, respectively,
and $W = 2 \sqrt{m_\Lambda^2 + (k^a)^2} = \sqrt{m_N^2 + (k^b)^2} + \sqrt{m_\Xi^2 + (k^b)^2}$,
$H_0^c = -\nabla^2/(2\mu^c)$ and $E_W^c = (k^c)^2/(2\mu^c)$
with $\mu^c$ being the reduced mass in the channel $c$.
The NBS wave function $\phi_W^c$ is defined in a similar matter to the 2N case.
The extension to the time-dependent HAL QCD method for coupled channel systems 
is straightforward~\cite{Aoki:2012tk}.
%

\subsection{Unified Contraction Algorithm (UCA)}
\label{subsec:UCA}

\begin{figure}[t]
\vspace*{-8mm}
\begin{minipage}{0.48\textwidth}
\begin{center}
\includegraphics[angle=0,width=0.80\textwidth]{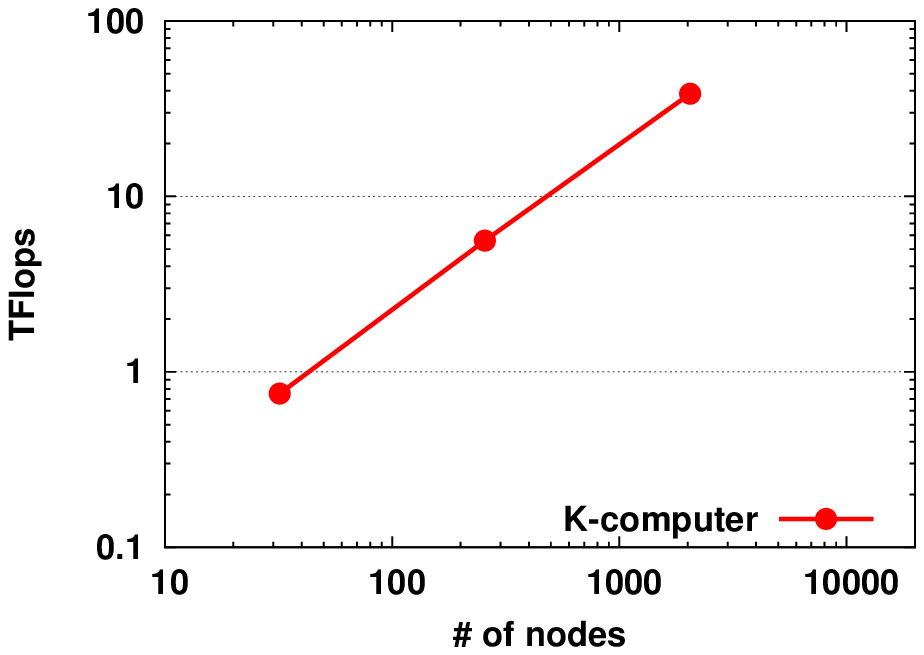}
\vspace*{-2mm}
\caption{
\label{fig:flops}
Weak scaling of the total flops performance 
for 
calculations of NBS correlators.
Measured on K computer with excluding the I/O part.
}
\end{center}
\end{minipage}
\hfill
\vspace*{-2mm}
\begin{minipage}{0.48\textwidth}
\begin{center}
\vspace*{-5mm}
\includegraphics[angle=0,width=0.80\textwidth]{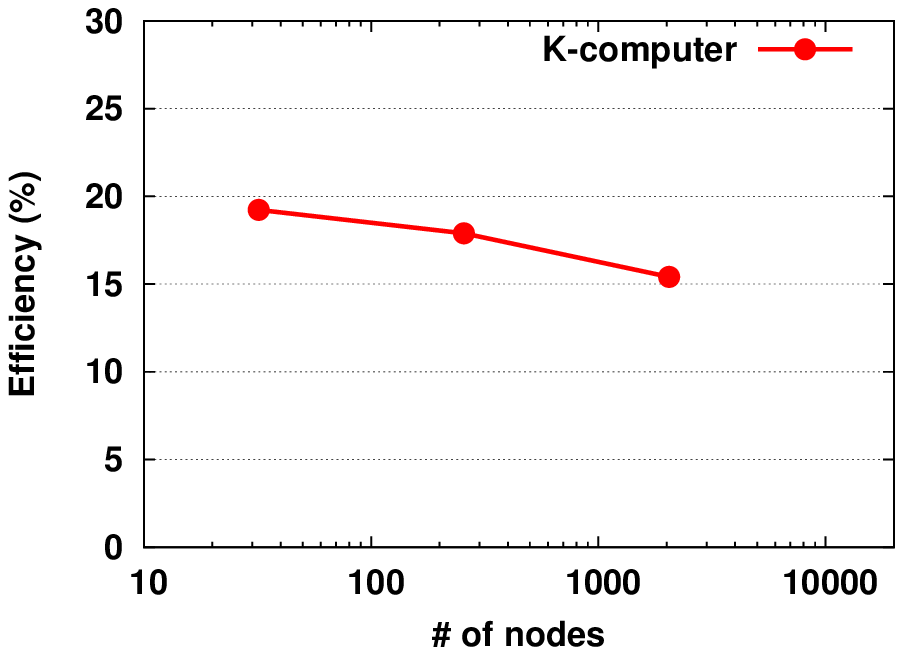}
\vspace*{-2mm}
\caption{
\label{fig:efficiency}
Same as Fig.~\protect\ref{fig:flops}, but for the performance efficiency.
}
\end{center}
\end{minipage}
\end{figure}

In lattice QCD studies of multi-baryon systems,
significant amount of computational resources
are required for the calculation of correlation functions.
The reasons are that 
(i) the number of 
Wick contractions
grows  
factorially with mass number $A$, and
(ii) the number of 
color/spinor contractions grows exponentially for larger $A$.
In the case of the HAL QCD method, additional resources are necessary since
(iii) each NBS correlation function has spacial degrees of freedom ($L^3$).
Note also that there are many channels (correlation functions) to be calculated
in order to obtain nuclear and hyperon forces.
Specifically, in $N_f = 2+1$ flavor space, there exist 52 channels in particle base
in the case of baryon forces between two-octet baryons 
in parity-even channel.

For (iii), we can reduce the computational cost by considering 
the ``baryon-block'' in momentum space together with FFT technique.
(A brief explanation is given as ``block algorithm'' in Ref.~\cite{Doi:2012xd}.)
While (i) and (ii) have been significant bottlenecks in lattice studies
for multi-baryon systems, 
we recently develop a novel algorithm
called the unified contraction algorithm (UCA).
This algorithm unifies the two contractions (i) and (ii) in a systematic way,
and significantly reduces the computational cost~\cite{Doi:2012xd}.
We implement UCA 
with the use of OpenMP + MPI hybrid parallelism
in our computational code for the NBS correlators,
called ``Hadron-Force code.''
In Figs.~\ref{fig:flops} and ~\ref{fig:efficiency},
we show the weak scaling of the total flops performance and its efficiency, respectively,
for the Hadron-Force code excluding the I/O part.
Good weak scaling with the efficiency of 15--20 \% is achieved 
for 32--2048 nodes ($\times$ 8 cores/node) on K computer.
Together with the efficient solver for quark propagators~\cite{Boku:2012zi},
the total measurement calculation (solver and Hadron-Force code as well as I/O)
achieves $\sim$ 17\% efficiency, or $\sim$ 45 TFlops sustained on 2048 nodes of K computer.

\section{Lattice QCD setup}
\vspace*{-2mm}
\label{sec:setup}

$N_f = 2+1$ gauge configurations are generated with
the Iwasaki gauge action at $\beta = 1.82$ and 
nonperturbatively ${\cal O}(a)$-improved Wilson quark action with $c_{sw} = 1.11$~\cite{Taniguchi:2012kk}
on the $96^4$ lattice.
We employ hopping parameters $(\kappa_{ud}, \kappa_s) = (0.126117, 0.124790)$
with APE stout smearing with $\alpha = 0.1$, $n_{\rm stout} = 6$
and the periodic boundary condition is used for all four directions.
Domain-Decomposed HMC (DDHMC) is used for ud quarks and 
UV-filtered Polynomial HMC (UVPHMC) is used for s quark.
Additional techniques such as even-odd preconditioning, 
mass preconditioning and multi-time scale integration are employed.
Using K computer, about 2000 trajectories are generated after the thermalization,
and preliminary studies show that $a^{-1} \simeq 2.33$ GeV ($a \simeq 0.085$ fm)
and $m_\pi \simeq 146$ MeV, $m_K \simeq 525$ MeV~\cite{Ukita:lat2015}.
This lattice setup enables us to study baryon forces with
physically light quark masses on the large lattice volume of $(8.2 {\rm fm})^4$.
For further details on the gauge configuration generation,
see Ref.~\cite{Ukita:lat2015}.

The measurements of NBS correlators are performed at the unitary point.
We calculate all 52 channels relevant to two-octet baryon forces in parity-even channel.
We employ wall quark source with Coulomb gauge fixing
(which is performed after the stout smearing).
The periodic boundary condition is used for spacial directions,
while the Dirichlet boundary condition is used for temporal direction
at $t-t_0 = 47, 48$ in order to avoid the wrapping-around artifact, 
and forward and backward propagations are averaged to reduce the statistical fluctuations.
We pick 1 configuration per each 10 trajectories,
and we make a use of the rotation symmetry to increase the statistics.
The total statistics used in this report amounts to
203 configurations $\times$ 4 rotations $\times$ 20 wall sources.

\begin{figure}[t]
\begin{minipage}{0.48\textwidth}
\begin{center}
\vspace*{-8mm}
\includegraphics[angle=0,width=0.85\textwidth]{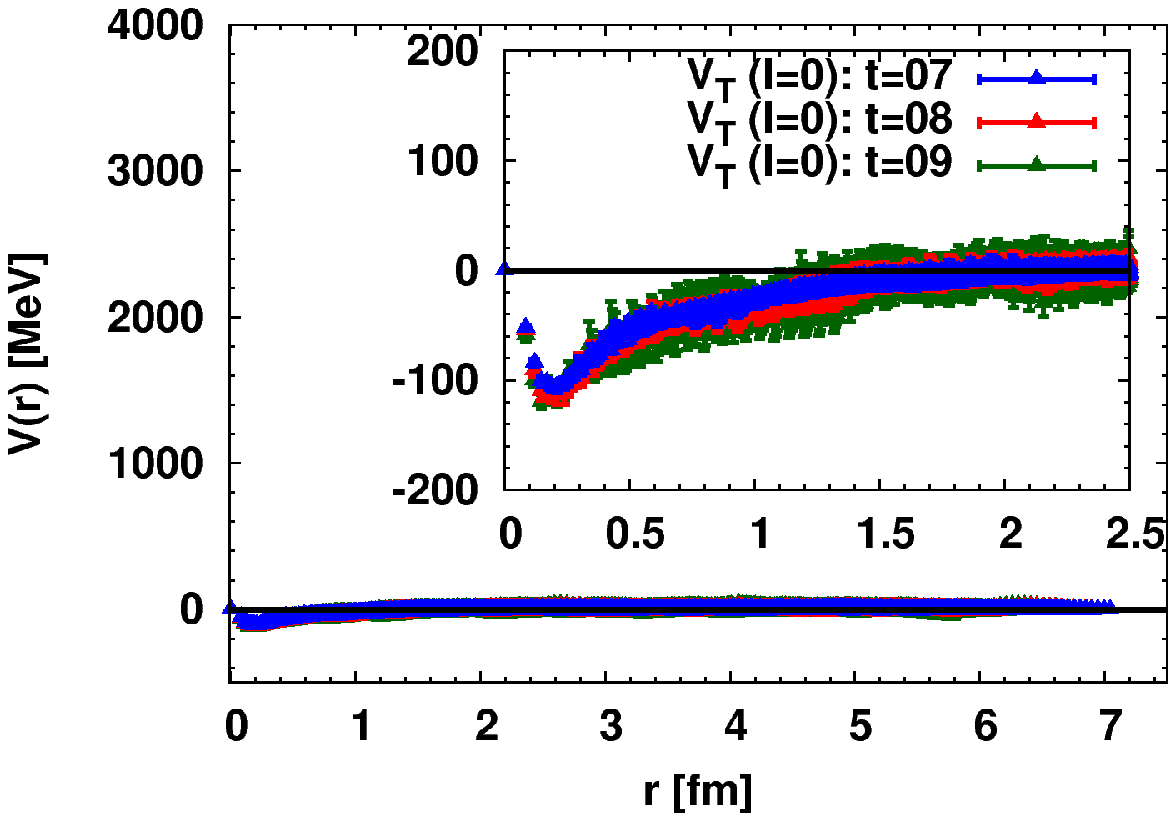}
\vspace*{-2mm}
\caption{
\label{fig:pot:NN:3S1:ten}
Nuclear tensor force $V_T(r)$ in $^3S_1$-$^3D_1$ $(I=0)$ channel
obtained at $t = 7, 8, 9$.
}
\end{center}
\end{minipage}
\hfill
\vspace*{-2mm}
\begin{minipage}{0.48\textwidth}
\begin{center}
\vspace*{-8mm}
\includegraphics[angle=0,width=0.85\textwidth]{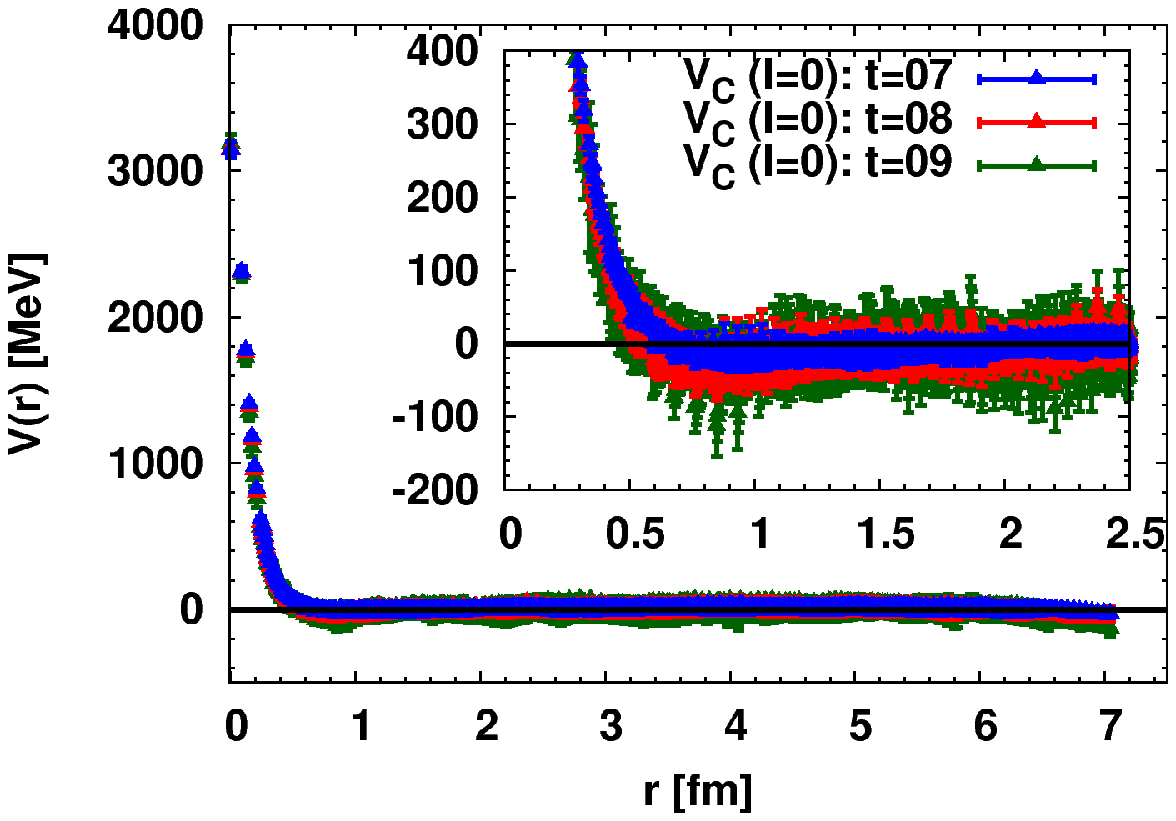}
\vspace*{-2mm}
\caption{
\label{fig:pot:NN:3S1:cen}
Nuclear central force $V_C(r)$ in $^3S_1$-$^3D_1$ $(I=0)$ channel
obtained at $t = 7, 8, 9$.
}
\end{center}
\end{minipage}
\end{figure}

\vspace*{-3mm}
\section{Results}
\vspace*{-2mm}
\label{sec:results}

Baryon forces are determined in the time-dependent HAL QCD method
in $^3S_1$-$^3D_1$ and $^1S_0$ channels.
We perform the velocity expansion~\cite{Aoki:2012tk} in terms of 
the non-locality of potentials,
and obtain the leading order potentials, i.e., central and tensor forces.
In this report, we present the results for nuclear forces:
See Refs.~\cite{Ishii:lat2015} for hyperon forces.
In this preliminary analysis shown below, 
the term which corresponds to the relativistic effects
($\partial^2 / \partial t^2$-term in Eq.~(\ref{eq:Sch_2N:tdep}))
is neglected.


Let us first consider the $^3S_1$-$^3D_1$ (iso-singlet) channel.
The $S$-wave and $D$-wave components in NBS correlators are extracted by 
${\cal P}$ and $(1-{\cal P})$-projection, respectively,
where ${\cal P}$-projection denotes the $A_1^+$ projection in the cubic group
and the higher partial wave mixings ($l \ge 4$) are neglected.
By solving the coupled channel Schr\"odinger equation, we obtain the tensor and central forces.
In Fig.~\ref{fig:pot:NN:3S1:ten}, we show the tensor force $V_T(r)$ obtained at $t = 7, 8, 9$.
Remarkably, it is clearly visible 
that $V_T(r) < 0$ with the long-range tail,
qualitatively in accordance with phenomenological potentials and/or the 
structure of the one-pion-exchange potential.
Compared to the lattice tensor forces obtained with heavier quark masses,
the range of interaction is found to be longer.
Since it is the tensor force which plays the most crucial role in 
the binding of deuteron,
this is a very 
intriguing
result.
For more quantitative studies, 
it is desirable to take larger $t$ by increasing the statistics, which is currently underway.

In Fig.~\ref{fig:pot:NN:3S1:cen}, we show the central force $V_C(r)$ in $^3S_1$-$^3D_1$ channel obtained at $t = 7, 8, 9$.
In this case, the potentials suffer from much larger statistical fluctuations.
However, the repulsive core at short-range is clearly obtained
and it is also encouraging that mid- and long-range attraction tends to appear as 
we take larger $t$.
Certainly, a study with larger $t$ with larger statistics is desirable.

\begin{figure}[t]
\begin{center}
\vspace*{-9mm}
\includegraphics[angle=0,width=0.432\textwidth]{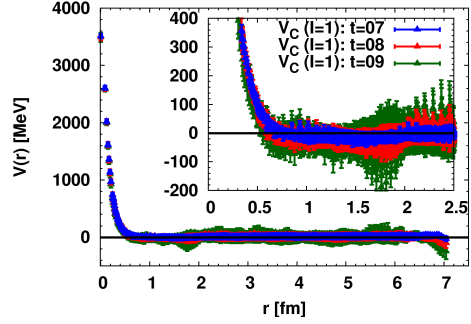}
\vspace*{-3mm}
\caption{
\label{fig:pot:NN:1S0:cen}
Nuclear central force $V_C(r)$ in $^1S_0$ $(I=1)$ channel
obtained at $t = 7, 8, 9$.
}
\vspace*{-6mm}
\end{center}
\end{figure}


We then consider the $^1S_0$ (iso-triplet) channel.
The $S$-wave components are extracted by ${\cal P}$-projection
and the central force is obtained by solving the single channel Schr\"odinger equation.
Shown in Fig.~\ref{fig:pot:NN:1S0:cen} is the obtained central force $V_C(r)$ at $t = 7, 8, 9$.
As was the case for the central force in $^3S_1$-$^3D_1$ channel,
the results suffer from large statistical fluctuations.
Yet, the repulsive core at short-range is observed
and the tendency is seen that mid- and long-range attraction emerges as we take larger $t$.
Investigations with larger $t$ and larger statistics are under progress.

\vspace*{-3mm}
\section{Summary}
\vspace*{-2mm}
\label{sec:summary}

We have presented the first lattice QCD studies 
for baryon interactions 
which employ 
physically light quark masses.
$N_f = 2+1$ gauge configurations have been generated with
the Iwasaki gauge action and 
nonperturbatively ${\cal O}(a)$-improved Wilson quark action with stout smearing
on the lattice of $(96 a)^4 \simeq (8.2 {\rm fm})^4$ with $a \simeq 0.085$ fm,
where 
$m_\pi \simeq 146$ MeV and $m_K \simeq 525$ MeV.
Baryon forces have been calculated from 
NBS 
correlation functions
in the HAL QCD method.

We have given a general overview of this project
and presented three developments crucial for this study:
(1) time-dependent HAL QCD method, by which notorious ground state saturation can be avoided
(2) coupled channel HAL QCD formalism above inelastic threshold,
which is particularly useful for hyperon forces
and
(3) unified contraction algorithm for the efficient computation.

Preliminary numerical results for the two-nucleon central and tensor forces in $^3S_1$-$^3D_1$ coupled channel
and
the central force in $^1S_0$ channel
have been shown.
In particular, we have observed a clear signal for the tensor force.
Together with the results for hyperon forces~\cite{Ishii:lat2015}
and with increased statistics available soon,
we expect to mark a significant milestone which bridges
particle physics and nuclear physics as well as astrophysics.

\vspace*{-2mm}
\section*{Acknowledgments}
\vspace*{-3mm}

The lattice QCD calculations have been performed on the K computer at RIKEN, AICS
(Nos. hp120281, hp130023, hp140209, hp150223),
HOKUSAI FX100 computer at RIKEN, Wako (No. G15023)
and HA-PACS at University of Tsukuba (Nos. 14a-20, 15a-30).
We thank ILDG/JLDG~\cite{conf:ildg/jldg}
which serves as an essential infrastructure in this study.
This work is supported in part by 
MEXT Grant-in-Aid for Scientific Research (15K17667, 25287046, 26400281),
and SPIRE (Strategic Program for Innovative REsearch) Field 5 project.
We thank all collaborators in this project.



\vspace*{-2mm}
\begin{thebibliography}{99}

\bibitem{Yukawa:1935xg}
  H.~Yukawa,
  Proc.\ Phys.\ Math.\ Soc.\ Jap.\  {\bf 17} (1935) 48
   [Prog.\ Theor.\ Phys.\ Suppl.\  {\bf 1} 1].


\bibitem{Luscher:1990ux}
  M.~\Luscher,
  Commun.\ Math.\ Phys.\  {\bf 105} (1986) 153;
%
  {\it ibid.},
  Nucl.\ Phys.\  B {\bf 354} (1991) 531.



\bibitem{Yamazaki:2011nd}
  T.~Yamazaki {\it et al.} [PACS-CS Collaboration],
  Phys.\ Rev.\ D {\bf 84} (2011) 054506
  [arXiv:1105.1418 [hep-lat]];
%
  T.~Yamazaki, K.~i.~Ishikawa, Y.~Kuramashi and A.~Ukawa,
  Phys.\ Rev.\ D {\bf 86} (2012) 074514
  [arXiv:1207.4277 [hep-lat]];
%
  {\it ibid.},
  Phys.\ Rev.\ D {\bf 92} (2015) 1,  014501
  [arXiv:1502.04182 [hep-lat]].



\bibitem{Beane:2011iw}
%
  S.~R.~Beane {\it et al.} [NPLQCD Collaboration],
  Phys.\ Rev.\ D {\bf 85} (2012) 054511
  [arXiv:1109.2889 [hep-lat]];
%
  S.~R.~Beane {\it et al.} [NPLQCD Collaboration],
  Phys.\ Rev.\ D {\bf 87} (2013) 3,  034506
  [arXiv:1206.5219 [hep-lat]];
%
  S.~R.~Beane {\it et al.} [NPLQCD Collaboration],
  Phys.\ Rev.\ C {\bf 88} (2013) 2,  024003
  [arXiv:1301.5790 [hep-lat]];
%
  K.~Orginos {\it et al.} [NPLQCD Collaboration],
  arXiv:1508.07583 [hep-lat].


\bibitem{Berkowitz:2015eaa}
  E.~Berkowitz {\it et al.} [CalLat Collaboration],
  arXiv:1508.00886 [hep-lat].


\bibitem{Ishii:2006ec}
  N.~Ishii, S.~Aoki and T.~Hatsuda,
  Phys.\ Rev.\ Lett.\  {\bf 99} (2007) 022001
  [nucl-th/0611096];
%
  S.~Aoki, T.~Hatsuda and N.~Ishii,
  Prog.\ Theor.\ Phys.\  {\bf 123} (2010) 89
  [arXiv:0909.5585 [hep-lat]].

\bibitem{HALQCD:2012aa}
  N.~Ishii {\it et al.} [HAL QCD Collaboration],
  Phys.\ Lett.\ B {\bf 712} (2012) 437
  [arXiv:1203.3642 [hep-lat]].

\bibitem{Aoki:2012tk}
  Reviewed in
  S.~Aoki {\it et al.} [HAL QCD Collaboration],
  Prog. Theor. Exp. Phys. {\bf 2012} (2012) 01A105
  [arXiv:1206.5088 [hep-lat]].

\bibitem{HAL:recent}
  F.~Etminan {\it et al.} [HAL QCD Collaboration],
  Nucl.\ Phys.\ A {\bf 928} (2014) 89
  [arXiv:1403.7284 [hep-lat]];
%
  M.~Yamada {\it et al.} [HAL QCD Collaboration],
  PTEP {\bf 2015} (2015) 7,  071B01
  [arXiv:1503.03189 [hep-lat]];
%
  K.~Sasaki {\it et al.} [HAL QCD Collaboration],
  PTEP {\bf 2015} (2015) 113B01 
  arXiv:1504.01717 [hep-lat];
%
  T.~Doi {\it et al.} [HAL QCD Collaboration],
  Prog.\ Theor.\ Phys.\  {\bf 127} (2012) 723
  [arXiv:1106.2276 [hep-lat]].

\bibitem{Inoue:2013nfe}
  T.~Inoue {\it et al.} [HAL QCD Collaboration],
  Phys.\ Rev.\ Lett.\  {\bf 111} (2013) 11,  112503
  [arXiv:1307.0299 [hep-lat]];
%
  T.~Inoue {\it et al.} [HAL QCD Collaboration],
  Phys.\ Rev.\ C {\bf 91} (2015) 1,  011001
  [arXiv:1408.4892 [hep-lat]].




\bibitem{Ishii:lat2015}
  N.~Ishii {\it et al.},
  in these proceedings;
%
  K.~Sasaki {\it et al.},
  in these proceedings;
%
  H.~Nemura {\it et al.},
  in these proceedings.

\bibitem{Ukita:lat2015}
  N.~Ukita {\it et al.},
  in these proceedings.


\bibitem{Lin:2001ek}
  C.~J.~D.~Lin {\it et al.}, 
  Nucl.\ Phys.\  B {\bf 619} (2001) 467
  [arXiv:hep-lat/0104006].


\bibitem{Kurth:2013tua}
  T.~Kurth, N.~Ishii, T.~Doi, S.~Aoki and T.~Hatsuda,
  JHEP {\bf 1312} (2013) 015
  [arXiv:1305.4462 [hep-lat]].

\bibitem{Iritani:2015dhu}
  T.~Iritani [HAL QCD Collaboration],
  in these proceedings,
  arXiv:1511.05246 [hep-lat].






\bibitem{Doi:2012xd}
  T.~Doi and M.~G.~Endres,
  Comput.\ Phys.\ Commun.\  {\bf 184} (2013) 117
  [arXiv:1205.0585 [hep-lat]].

\bibitem{Boku:2012zi}
  T.~Boku {\it et al.},
  PoS LATTICE {\bf 2012} (2012) 188
  [arXiv:1210.7398 [hep-lat]].


\bibitem{Taniguchi:2012kk}
  Y.~Taniguchi,
  PoS LATTICE {\bf 2012} (2012) 236
  [arXiv:1303.0104 [hep-lat]].


\bibitem{conf:ildg/jldg}
  \verb|"http://www.lqcd.org/ildg"|,
  \verb|"http://www.jldg.org"|

\end{thebibliography}
\end{document}